\documentclass[10pt,conference]{IEEEtran}
\IEEEoverridecommandlockouts
\usepackage{cite}
\usepackage{tabularx}
\usepackage{amsmath,amssymb,amsfonts}
\usepackage{algorithmic}
\usepackage{graphicx}
\usepackage{textcomp}
\usepackage{xcolor}
\usepackage{tabularx}
\usepackage{multirow}
\usepackage{float}
\usepackage{caption}
\usepackage{url}
\usepackage{tcolorbox}
\usepackage{array}
\usepackage{subcaption}
\usepackage{arydshln} 
\usepackage{listings} 
\usepackage{courier}  
\usepackage{amsmath}  
\usepackage{anyfontsize}
\usepackage{array} 
\usepackage{url}
\usepackage{booktabs}
\usepackage{graphicx}
\usepackage[hidelinks]{hyperref}  

\def\BibTeX{{\rm B\kern-.05em{\sc i\kern-.025em b}\kern-.08em
    T\kern-.1667em\lower.7ex\hbox{E}\kern-.125emX}}
\begin{document}

\title{When More Retrieval Hurts: Retrieval-Augmented Code Review Generation}
\author{\IEEEauthorblockN{1\textsuperscript{st} Qianru Meng}
\IEEEauthorblockA{\textit{LIACS} \\
\textit{Leiden University}\\
Leiden, Netherlands \\
mengqr@vuw.leidenuniv.nl}
\and
\IEEEauthorblockN{2\textsuperscript{nd} Xiao Zhang}
\IEEEauthorblockA{\textit{CLCG} \\
\textit{University of Groningen}\\
Groningen, Netherlands \\
xiao.zhang@rug.nl}
\and
\IEEEauthorblockN{3\textsuperscript{rd} Zhaochun Ren}
\IEEEauthorblockA{\textit{LIACS} \\
\textit{Leiden University}\\
Leiden, Netherlands \\
z.ren@liacs.leidenuniv.nl}
\and
\IEEEauthorblockN{4\textsuperscript{th} Joost Visser}
\IEEEauthorblockA{\textit{LIACS} \\
\textit{Leiden University}\\
Leiden, Netherlands \\
j.m.w.visser@liacs.leidenuniv.nl}
}

\maketitle

\begin{abstract}
Code review generation can reduce developer effort by producing concise, reviewer-style feedback for a given code snippet or code change.
However, generation-only models often produce generic or off-point reviews, while retrieval-only methods struggle to adapt well to new contexts.
In this paper, we view retrieval augmentation for code review as \emph{retrieval-augmented in-context learning}: retrieved historical reviews are placed in the input context as examples that guide the model’s output.
Based on this view, we propose \textbf{RARe} (Retrieval-Augmented code Reviewer), a framework that retrieves relevant historical reviews from a corpus and conditions a large language model on the retrieved in-context examples.
Experiments on two public benchmarks show that RARe outperforms strong baselines and reaches BLEU-4 scores of 12.32 and 12.96.
A key finding is that \textbf{more retrieval can hurt}: using only the top-1 retrieved example works best, while adding more retrieved items can degrade performance due to redundancy and conflicting cues under limited context budgets.
Human evaluation and interpretability analysis further support that retrieval-augmented generation reduces generic outputs and improves review focus.
\end{abstract}


\section{Introduction}
\label{sec:introduction}

Code review is a core practice for improving software quality, but it is time-consuming and knowledge-intensive.
Automatically generating code reviews is therefore attractive for (i) assisting reviewers with candidate feedback and (ii) reducing developer turnaround time by providing immediate critique.
Prior work spans rule-based static analysis and repository mining~\cite{palvannan2023suggestion,balachandran2013reducing,denney2009generating,klinik2021personal,chatley2018diggit},
neural generation approaches~\cite{gupta2018intelligent,siow2020core,tufano2022using,li2022crer,li2022auger,lu2023llama},
and retrieval-based methods that reuse historical reviews for similar code~\cite{hong2022commentfinder}.
Retrieval is efficient and can recover project- or domain-specific tokens, but retrieval-only systems cannot flexibly adapt reviews to new contexts.

Although both retrieval and generation can work, each has a characteristic failure mode in code review.
Retrieval-only approaches are bounded by the corpus and may return lexically similar yet intent-misaligned reviews.
Generation-only approaches can produce novel feedback, but LLM outputs often become overly generic, summary-like, or off-point in review settings~\cite{li2022auger}.
Fine-tuning can improve task alignment, yet may overfit to limited data and still miss the intended review focus.

This paper takes a task-specific retrieval-augmented generation perspective for code review.
We view retrieval augmentation as \textbf{retrieval-augmented in-context learning}, where retrieved historical reviews are placed in the input context as \emph{in-context examples}.
These examples provide signals about (i) \emph{review style} (short, actionable, reviewer-like) and (ii) \emph{review focus} (what to critique in the code).
Crucially, historical reviews encode both intent and tone; naively adding multiple retrieved items may introduce redundant or conflicting cues.
Therefore, \textbf{deciding how many retrieved examples to include} becomes central.

We propose \textbf{RARe}, a retrieval-augmented framework for code review generation that combines a learned retriever with a large language model.
RARe consists of (1) a retriever that finds relevant historical reviews from a corpus and (2) a generator that produces the final review conditioned on the target code and the retrieved in-context example(s).

Experiments on two benchmarks show that RARe improves over strong baselines, reaching BLEU-4 scores of 12.32 and 12.96.
Beyond automatic metrics, we conduct human evaluation and interpretability-based analysis to explain \emph{why} retrieval helps: retrieved examples reduce generic summaries and steer generation toward targeted, reviewer-like critiques.

\textbf{Contributions.} We make the following contributions:
\begin{itemize}
    \item We present RARe, a retrieval-augmented framework for code review generation that uses retrieved historical reviews as in-context examples.
    \item We provide a component-wise study of retrievers (NDR/GDR/DPR) and generators (Llama~3.1, Mistral-7B, CodeGemma-7B) under direct inference and LoRA fine-tuning.
    \item Our ablation studies show that \textbf{top-1 retrieval is best}: adding more retrieved examples can hurt, highlighting an important non-monotonic behavior for code review generation.
    \item We complement automatic metrics with human evaluation and interpretability-based evidence to explain how retrieval influences style and review focus.
    \item All code is available in an anonymous repository: \url{https://anonymous.4open.science/r/GAR-9EE2}.
\end{itemize}

\begin{figure*}[t]
  \centering
  \includegraphics[width=\linewidth]{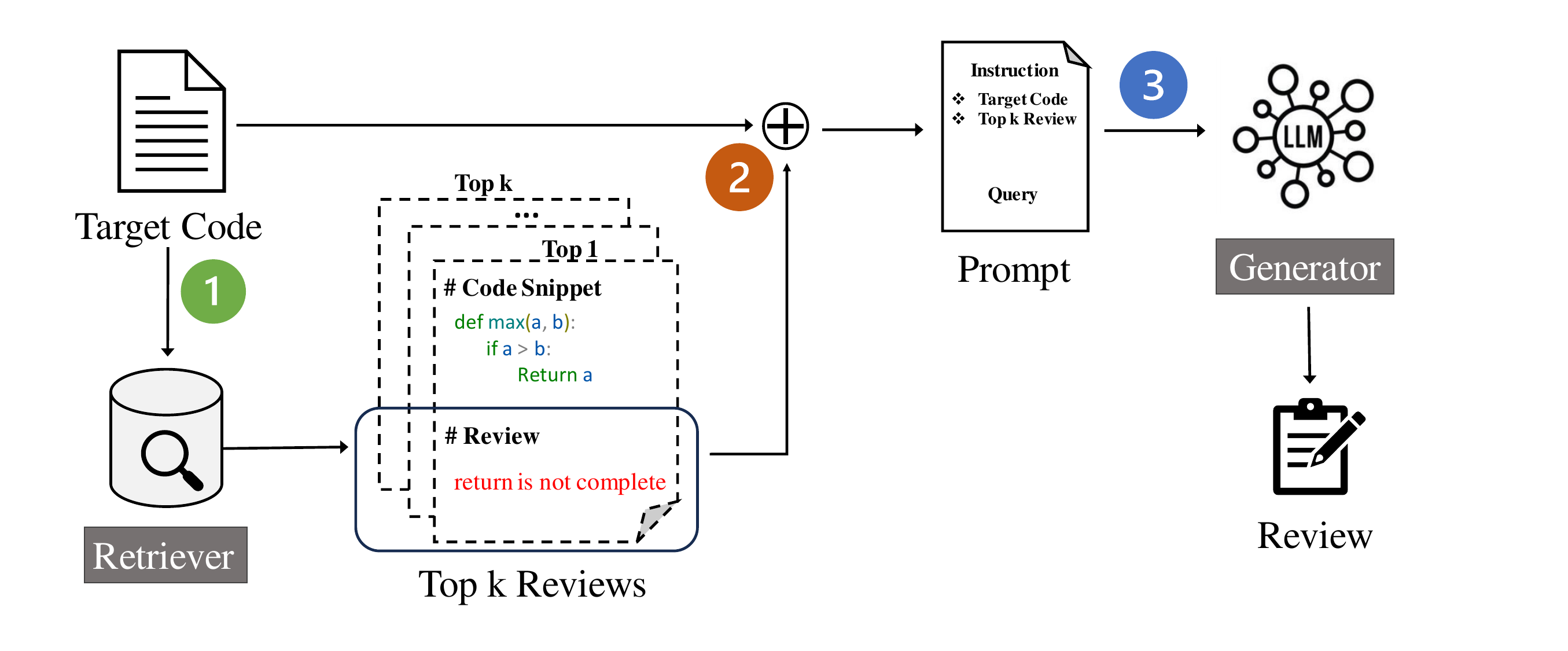}
  \caption{The overall architecture of RARe. Different colors distinguish the retrieval, generation, and RAG processes. Within the review of the target code, the text related to generation is marked in blue, while those related to retrieval are marked in green.}
  \label{fig:pipeline}
\end{figure*}

\section{Related Work}
\label{sec:related_work}

\subsection{Automation of Code Review Comment Generation}
Early studies used static analysis and hand-crafted rules to automate parts of review processes~\cite{palvannan2023suggestion, balachandran2013reducing, denney2009generating,klinik2021personal}.
For instance, Balachandran et al.~\cite{balachandran2013reducing} developed a review bot integrating multiple analyzers to suggest code modifications.
Palvannan et al.~\cite{palvannan2023suggestion} built a suggestion bot for GitHub using a Python analyzer.
Repository mining has also been used to derive validation rules from historical commits~\cite{chatley2018diggit}.
While efficient, rule-based approaches lack portability and struggle with open-ended, context-dependent review comments.

Neural approaches improved semantic understanding of code and reviews.
Gupta et al.~\cite{gupta2018intelligent} introduced DeepCodeReviewer to model relevance and recommend historical reviews.
Siow et al.~\cite{siow2020core} proposed CORE with attention-based LSTM to generate comments.
With transformer pre-training~\cite{NIPS2017_3f5ee243}, later work adopted seq2seq frameworks and task-specific objectives for review tasks~\cite{li2022auger,tufano2022using,li2022crer}.
More recently, LLMs have been fine-tuned for code review comment generation; Lu et al.~\cite{lu2023llama} studied tuning strategies and proposed LLaMA-Reviewer.

Retrieval-based review generation is a complementary paradigm: given a code snippet/change, retrieve similar historical items and reuse their reviews.
CommentFinder~\cite{hong2022commentfinder} represents code with bag-of-words~\cite{zhang2010bow} and combines cosine similarity with Gestalt Pattern Matching (GPM)~\cite{ratcliff1988gpm} for reranking.
Retrieval excels at efficiency and at copying low-frequency but important tokens, yet cannot reliably adapt to novel contexts.

\subsection{Retrieval-Augmented Generation}
Retrieval-augmented generation (RAG) was proposed in early neural settings~\cite{Lewis-2020-RAG,karpukhin2020dpr} and has been widely adopted for question answering and dialogue~\cite{guu2020retrieval,Lewis-2020-RAG,fabbri-etal-2020-template,li-etal-2023-unified,fan2021augmenting,king-flanigan-2023-diverse,cheng2024rag, zhang-etal-2025-retrieval}.
LLMs renewed broad interest in RAG because they can incorporate retrieved context through prompting~\cite{gao2024retrievalaugmented}.
In many LLM-based RAG pipelines, retrieved items are inserted into the prompt as in-context examples that influence generation.


RAG has also been used for code-related tasks, including code generation and summarization~\cite{parvez2021ragdpr}, code completion~\cite{lu2022reacc}, assertion generation and program repair~\cite{nashid2023retrieval}, comment generation~\cite{geng2024large}, and commit message generation~\cite{zhang2024rag}.
For code review generation, retrieval augmentation is particularly motivated because review intent and style are strongly reflected in historical reviews.
Recent work has explicitly explored retrieval augmentation for code review generation, e.g., retrieval-augmented code review comment generation~\cite{hong2025ragrcg} and LAURA~\cite{zhang2025laura}.
These studies show the promise of augmenting review generation with retrieved historical context, but they leave open how retrieval design choices affect generation quality across different evaluation settings.
In parallel, recent RAG research has emphasized \emph{selective retrieval and context compression} to mitigate redundancy when prompts are tight, including EXIT~\cite{hwang2025exit}, PISCO~\cite{louis2025pisco}, and SARA~\cite{jin2025sara}. 

Prior work has already explored retrieval augmentation for code review generation, but existing studies have mainly evaluated it on single-language corpora. In contrast, we provide a more comprehensive evaluation by considering both single-language and cross-language benchmarks. We further study code review generation from the perspective of \emph{retrieval-augmented in-context learning}, where retrieved reviews mainly serve as examples that shape review focus and wording. This perspective highlights an underexplored question in prior work: how many retrieved examples are actually useful.

\section{Methodology}
\label{sec:methodology}

RARe is a retrieval-augmented framework for code review generation.
Given a target code snippet or code change $x$, RARe retrieves relevant historical review(s) from a corpus and conditions an LLM on the target code together with the retrieved review example(s) to generate the final review $o$.
Figure~\ref{fig:pipeline} illustrates the pipeline.

\subsection{Problem Formulation}
Let $\mathcal{D}=\{(x_i,o_i)\}_{i=1}^{N}$ denote paired code $x$ and human-written review $o$.
Given a new input $x$, the goal is to generate a concise review $o$ that matches reviewer-style writing and focuses on relevant issues.

\begin{table}[t]
\centering
\small
\setlength{\tabcolsep}{3pt}
\renewcommand{\arraystretch}{1.05}
\caption{Prompt templates used in our experiments.}
\label{tab:prompt}
\begin{tabular}{p{\columnwidth}}
\toprule
\textbf{Direct inference.}\\
Your task is to write a concise code review for the given code snippet. Your output should only be a brief code review, no extra information.\\
\midrule
\textbf{RARe.}\\
Here is an example code review: \{retrieved review(s)\}.\\
Your task is to write a concise code review for the given code snippet. Your output should only be a brief code review, no extra information.\\
\bottomrule
\end{tabular}
\end{table}

RARe has three steps:
\begin{enumerate}
    \item \textbf{Retrieve:} obtain top-$k$ historical items $\{(x'_j,o'_j)\}_{j=1}^{k}$ relevant to $x$.
    \item \textbf{Construct context:} combine the instruction, retrieved review(s) $\{o'_j\}$, and target code $x$.
    \item \textbf{Generate:} use an LLM to produce the review $o$ from the augmented context.
\end{enumerate}

\begin{table*}[htbp]
\centering
\small
\caption{Comparison of three retrievers: NDR (Normal Dense Retrieval), GDR (GPM Dense Retrieval), DPR (Dense Passage Retrieval).}
\label{tab:retrieval_performance}
\begin{tabular}{l ccc ccc}
\toprule
Method & \multicolumn{3}{c}{CRer.} & \multicolumn{3}{c}{Tuf.} \\
\cmidrule(lr){2-4} \cmidrule(lr){5-7}
& BLEU-4 & ROUGE-L & METEOR & BLEU-4 & ROUGE-L & METEOR \\
\midrule
NDR & 9.53 & 5.65 & 5.54 & 12.06 & 8.45 & 8.17 \\
GDR & 9.55 & 5.66 & 5.55 & \textbf{12.33} & \textbf{8.67} & \textbf{8.43} \\
DPR & \textbf{9.71} & \textbf{5.70} & \textbf{5.60} & 11.80 & 7.98 & 7.74 \\
\bottomrule
\end{tabular}
\end{table*}

\begin{table*}[t]
\small
\centering
\caption{Performance of models on CRer. dataset and Tuf. dataset with and without Retrieval Augmentation (+RA).}
\label{tab:model_performance}
\begin{tabular}{l p{0.8cm} ccc ccc}
\toprule
\multicolumn{2}{l}{\multirow{2}{*}{Model}} & \multicolumn{3}{c}{CRer.} & \multicolumn{3}{c}{Tuf.} \\
\cmidrule(lr){3-5} \cmidrule(lr){6-8}
& & BLEU-4 & ROUGE-L & METEOR & BLEU-4 & ROUGE-L & METEOR \\
\midrule
\multirow{4}{*}{Llama-3.1-8B}
& DI & 5.84 & 3.18 & 3.18 & 5.12 & 2.55 & 2.67 \\
& +RA & \textbf{12.32} & \textbf{6.43} & \textbf{8.33} & \textbf{12.96} & \textbf{8.67} & \textbf{8.76} \\
& FT & 8.82 & 6.75 & 5.29 & 8.06 & 5.87 & 4.57 \\
& +RA & 9.05 & 6.90 & 5.42 & 9.19 & 7.00 & 5.66 \\
\midrule
\multirow{4}{*}{Mistral-7B}
& DI & 5.07 & 2.69 & 2.72 & 5.21 & 2.75 & 3.09 \\
& +RA & 11.15 & 5.75 & 6.87 & 9.74 & 7.69 & 7.58 \\
& FT & 9.07 & 6.56 & 5.32 & 8.42 & 6.21 & 4.91 \\
& +RA & 9.59 & 7.40 & 6.59 & 9.58 & 7.40 & 6.05 \\
\midrule
\multirow{4}{*}{CodeGemma-7B}
& DI & 3.34 & 1.53 & 1.40 & 5.27 & 3.02 & 2.95 \\
& +RA & 4.76 & 2.56 & 2.49 & 5.51 & 3.02 & 3.53 \\
& FT & 9.26 & 6.48 & 5.22 & 8.39 & 5.70 & 4.53 \\
& +RA & 9.33 & 6.51 & 5.25 & 9.44 & 6.88 & 5.60 \\
\bottomrule
\end{tabular}
\end{table*}

\subsection{Retriever}
\label{subsec:retriever}

The retriever ranks historical items by relevance to the target code $x$.
To avoid information leakage, the retrieval index is built only from the training split, and all validation/test instances retrieve exclusively from $\mathcal{D}_{\mathrm{train}}=\{(x_i,o_i)\}$.

For dense retrieval, given vectors $u$ and $v$, we use cosine similarity:
\begin{equation}
\mathrm{sim}(u,v)=\frac{u^\top v}{\|u\|\cdot \|v\|}.
\end{equation}
Candidates are ranked by similarity and the top-$k$ are returned.

We compare three retrieval methods: Normal Dense Retrieval (NDR), GPM Dense Retrieval (GDR), and Dense Passage Retrieval (DPR).

\subsubsection{Normal Dense Retrieval (NDR)}
NDR encodes code snippets into a shared embedding space with a single encoder $c(\cdot)$.
For target $x$ and candidate code $z$, relevance is
\begin{equation}
s_{\mathrm{NDR}}(x,z)=\mathrm{sim}\!\big(c(x),c(z)\big),
\end{equation}
where $c(\cdot)$ is implemented with a code encoder such as CodeBERT.
The paired review of the retrieved code is used as the example review.

\subsubsection{GPM Dense Retrieval (GDR)}
GDR is a two-stage method.
It first applies NDR to obtain a candidate pool, then reranks candidates using an order-sensitive token matching score based on Gestalt Pattern Matching (GPM)~\cite{ratcliff1988gpm}.
This is useful when token order is informative, especially in single-language corpora.

\subsubsection{Dense Passage Retrieval (DPR)}
DPR uses two encoders: a code encoder $c(\cdot)$ for input code and a review encoder $r(\cdot)$ for historical reviews.
It scores code--review relevance directly:
\begin{equation}
s_{\mathrm{DPR}}(x,o)=\mathrm{sim}\!\big(c(x),r(o)\big).
\end{equation}
DPR is trained contrastively~\cite{karpukhin2020dpr} to align matched code--review pairs and separate mismatched ones.
Compared with NDR, it is less dependent on code-to-code similarity and can be more robust in multi-language settings.

We vary the number of retrieved reviews as $k\in\{1,3,5\}$.

\subsection{Generator}
\label{subsec:generator}

We use decoder-only LLMs as generators.
Let $s$ denote the full input context, consisting of the instruction, retrieved review example(s), and target code $x$.
The generator models
\begin{equation}
p(o\mid s)=\prod_{i=1}^{n} p_{\theta}\!\left(o_i \mid s, o_{1:i-1}\right),
\end{equation}
where $o=(o_1,\dots,o_n)$ and $\theta$ denotes model parameters.

\subsection{Retrieval-Augmented Generation in RARe}
\label{subsec:rag_raRe}

Given input $x$, the retriever returns top-$k$ historical examples, which are prepended to the prompt as in-context examples.
These examples guide the model toward reviewer-like writing and relevant critique targets.

Let $\mathcal{R}_k(x)=\{o'_1,\dots,o'_k\}$ denote the retrieved review examples for $x$.
RARe generates
\begin{equation}
p_{\mathrm{RARe}}(o\mid x)=p_{\theta}\!\left(o \mid \mathrm{Prompt}(x,\mathcal{R}_k(x))\right),
\end{equation}
where $\mathrm{Prompt}(\cdot)$ formats the instruction, retrieved example(s), and target code into a single input sequence.
\section{Experimental Setup}
\label{sec:experiment}

\subsection{Research Questions}
\begin{itemize}
    \item \textbf{RQ1: How do retriever and generator choices affect RARe?}
    \item \textbf{RQ2: Does RARe outperform strong prior baselines?}
    \item \textbf{RQ3: How does the number of retrieved examples affect performance when prompts are limited?}
\end{itemize}

\subsection{Datasets}
\label{subsec:datasets}

We evaluate on CodeReviewer (CRer.)~\cite{li2022crer} and Tufano (Tuf.)~\cite{tufano2022using}.
CRer.\ contains GitHub code review instances spanning nine programming languages and represents code changes in a diff format; it has approximately 118K/10K/10K train/validation/test instances.
Tuf.\ is collected from GitHub and Gerrit, consists of Java instances, and provides code in a snippet format rather than diffs; it has approximately 134K/17K/17K instances.
We use the official splits released with each dataset and apply no additional preprocessing.

\subsection{Evaluation Metrics}
Following prior work~\cite{tufano2022using,hong2022commentfinder,li2022auger,li2022crer,lu2023llama}, we report BLEU-4, ROUGE-L, and METEOR.

\subsection{Baselines}
We compare against representative retrieval-only and generation-based baselines:
CommentFinder~\cite{hong2022commentfinder},
Tufano et al.~\cite{tufano2022using},
CodeReviewer~\cite{li2022crer},
AUGER~\cite{li2022auger},
LLaMA-Reviewer~\cite{lu2023llama},
and CodeT5~\cite{wang2021codet5}.

\subsection{Model Configurations and Training}
\textbf{Retrievers.}
We compare the performance of three retrievers (NDR, GDR and DPR).
Only DPR requires training and we train two encoders on the training data from the two datasets, following the well-adjusted hyperparameters provided by the original work \cite{parvez2021ragdpr}.
\begin{table*}[t]
\centering
\small
\caption{Overall comparison between baseline methods and RARe. $-$ indicates that the previous study did not provide the predictions. For RARe, the number of retrieved reviews are in brackets.}
\label{tab:overall_results}
\begin{tabular}{l ccc ccc}
\toprule
Method & \multicolumn{3}{c}{CRer.} & \multicolumn{3}{c}{Tuf.} \\
\cmidrule(lr){2-4} \cmidrule(lr){5-7}
& BLEU-4 & ROUGE-L & METEOR & BLEU-4 & ROUGE-L & METEOR \\
\midrule
Tufano et al.~\cite{tufano2022using} & - & - & - & 12.32 & 8.72 & 7.83 \\
CodeT5~\cite{wang2021codet5} & 7.34 & 7.41 & 5.86 & 7.10 & 6.61 & 5.13 \\
CodeReviewer~\cite{li2022crer} & 6.02 & 5.39 & 3.68 & - & - & - \\
CommentFinder~\cite{hong2022commentfinder} & 9.47 & 5.69 & 5.44 & 12.71 & 8.81 & 8.61 \\
AUGER~\cite{li2022auger} & 8.09 & 6.50 & 4.74 & 7.76 & 5.77 & 4.36 \\
LLaMA-Reviewer~\cite{lu2023llama} & 8.23 & 6.12 & 5.34 & 7.85 & 5.82 & 4.38 \\
\midrule
RARe (random) & 11.29 & 6.41 & 7.97 & 10.88 & 6.66 & 7.82 \\
\textbf{RARe (top1)} & \textbf{12.32} & \textbf{6.43} & \textbf{8.33} & \textbf{12.96} & \textbf{8.67} & \textbf{8.76} \\
RARe (top3) & 11.76 & 6.14 & 8.14 & 11.74 & 6.89 & 8.00 \\
RARe (top5) & 10.81 & 5.89 & 7.76 & 10.80 & 6.47 & 8.23 \\
\bottomrule
\end{tabular}
\end{table*}
\textbf{Generators.}
We evaluate three open-source LLMs: Llama-3.1-8B\footnote{\url{https://huggingface.co/meta-llama/Meta-Llama-3.1-8B-Instruct}}, Mistral-7B\footnote{\url{https://huggingface.co/mistralai/Mistral-7B-Instruct-v0.3}}, and CodeGemma-7B-7B\footnote{\url{https://huggingface.co/google/codegemma-7b-it}}
We use the same instruction prompt for all models; RARe additionally prepends retrieved review example(s).
Full prompt templates are provided in Table ~\ref{tab:prompt}.

We consider direct inference (DI) and LoRA fine-tuning (FT), and their retrieval-augmented variants where retrieved review example(s) are prepended to the prompt.
DI requires no parameter updates, whereas FT trains LoRA adapters on the training split (base model weights remain frozen).
We train LoRA for 5 epochs with learning rate $10^{-4}$ and batch size 16 on 4 NVIDIA A100 GPUs, and report averages over three runs.

\section{Results and Analysis}
\label{sec:results}

We report results under retrieval-augmented in-context learning, where retrieved historical reviews are included in the input context as examples that guide generation.
This section examines (i) how retriever and generator choices affect performance, (ii) how RARe compares with prior methods, and (iii) when adding more retrieved examples becomes detrimental.
We then present a case study, human evaluation, and interpretability evidence to complement the aggregate metrics.

\subsection{Comparison of retrieves}
\label{sec:retriever_choice}

As shown in Table~\ref{tab:retrieval_performance}. We focus on top-1 performance of three retrievers. On CRer., DPR achieves the best results (BLEU-4 9.71, ROUGE-L 5.70, METEOR 5.60), while on Tuf., GDR performs best (BLEU-4 12.33, ROUGE-L 8.67, METEOR 8.43).

Overall, GDR improves over NDR due to its GPM-based reranking, which better prioritizes relevant items. For the single-language Tuf. dataset, code-only similarity is often sufficient, making NDR/GDR competitive. In contrast, CRer.\ spans multiple languages, where code-only similarity is less reliable; DPR is more robust because it directly models code--review alignment. Therefore, we use DPR for CRer.\ and GDR for Tuf.\ in subsequent experiments.

\subsection{Comparison of generators}
\label{sec:effect_of_retrieval}
As introduced in Section~\ref{sec:experiment}, we compare three LLMs (Llama-3.1-8B, Mistral-7B and CodeGemma-7B) with employing four generation approaches: direct inference (DI), fine-tuning (FT), retrieval-augmented direct inference, and retrieval-augmented fine-tuning. The results are shown in Table~\ref{tab:model_performance}.  

Without retrieval augmentation, fine-tuning consistently improves direct inference across all three models and both datasets, indicating that task-specific training helps models better match the target review distribution. Llama~3.1 performs strongest under direct inference, while after fine-tuning the performance gap among the three models becomes smaller, suggesting that additional supervision reduces differences stemming from pre-training and architecture.

Retrieval augmentation provides the largest gains in direct inference. In particular, Llama~3.1 achieves the best overall results with retrieval augmentation on both datasets (CRer.: BLEU-4 12.32; Tuf.: BLEU-4 12.96). Retrieval also improves fine-tuned models, but the gains are smaller; among the three models, Mistral-7B shows the most consistent improvement after fine-tuning, while CodeGemma-7B benefits only marginally from retrieval in our setting. It may attribute to model's relatively weaker ability to understand prompts.

\begin{table*}[t]
\centering
\small
\caption{An example from Tuf.\ comparing RARe (CodeGemma-7B) with baselines. We omit some particularly long reviews.}
\label{tab:case_study_compare}
\renewcommand{\arraystretch}{1.15}

\begin{tabularx}{\linewidth}{
  >{\raggedright\arraybackslash}p{2.5cm}
  >{\raggedright\arraybackslash}X
  >{\raggedleft\arraybackslash}p{1.2cm}
}
\toprule
\textbf{System} & \textbf{Code/Review} & \textbf{BLEU-4} \\
\midrule

Code snippet &
\parbox[t]{\linewidth}{%
\ttfamily\footnotesize
public void onClose(Client client) \{\\
\quad synchronized(clients) \{\\
\quad\quad clients.remove(client);\\
\quad \}\\
\}%
} & \\

Ground Truth & This synchronization seems unnecessary? & \\
\hdashline
CodeT5 & Should this be a WARN instead? & 0.0839 \\
CommentFinder & Off topic: I've been writing this as ``client == null''... & 0.0442 \\
Tufano & I think you should have done something like: if (clients != null) \{ clients.remove(client); \} & 0.0255 \\
AUGER & Remove client from the interface but keep it on the i imp & 0.0530 \\
\hdashline
Retrieved Example & I don't see any other method that this synchronization would avoid overlapping with. & 0.2741 \\
DI & The code snippet synchronizes on the \texttt{clients} object before removing a client... & 0.1154 \\
DI+RA & The synchronized block is not necessary... & 0.1228 \\
FT & \texttt{synchronized (clients) \{ clients.remove(client); \}} & 0.2328 \\
FT+RA & this can be simplified to \texttt{clients.remove(client)} & 0.0895 \\
\bottomrule
\end{tabularx}
\renewcommand{\arraystretch}{1}
\end{table*}

\subsection{Comparison with prior methods}
\label{sec:comparison_prior}

Based on the previous comparison of retrievers and generators, we selected the best-performing retriever and generator for RARe to compare against the baseline methods. Specifically, for the CRer. dataset, we use the DPR retriever, while for the Tuf. dataset, we use the GDR retriever. And Llama-3.1-8B serves as the generator for both datasets. We conducted performance comparisons among RARe and baseline methods on both datasets, as shown in Table~\ref{tab:overall_results}. 
 
Compared to baselines, RARe achieves the best performance across both datasets in three metrics. Especially on the CRer. dataset, RARe significantly outperforms Commentfinder by 30\% in BLEU-4 and 53\% in METEOR, respectively. On the Tufano dataset, RARe shows a modest improvement, surpassing CommentFinder by 2\% in both BLEU-4 and METEOR. 

\subsection{How many retrieved examples should be used}
\label{sec:how_many}

Table~\ref{tab:overall_results} also varies the number of retrieved examples included in the input context.
A consistent non-monotonic trend appears across both datasets: top-1 performs best, while top-3 and top-5 degrade performance.
This suggests that retrieved reviews are not interchangeable evidence snippets: they encode critique direction and writing preferences.
When multiple retrieved reviews contain redundant or mismatched cues, the generator may mix incompatible signals within a limited context budget, leading to worse outputs.


\subsection{Case study}
\label{sec:case_human_inseq}

Automatic metrics provide aggregate comparisons, but they do not reveal how retrieval changes the generated review in individual instances.
We therefore provide (i) a qualitative case study, (ii) a human evaluation of usefulness, and (iii) interpretability evidence that the model relies on retrieved cues.

\subsubsection{Qualitative case study}
\label{sec:qual_case}

Table~\ref{tab:case_study_compare} shows a randomly selected example from Tuf.\ and compares multiple systems.
Two behaviors are particularly relevant to interpreting retrieval augmentation.

First, direct inference often produces descriptive text that explains what the code does.
Such outputs can be reasonable summaries but are frequently misaligned with the review convention in these datasets, where reviewers write short critiques and suggestions.
With retrieval augmentation, the output is more likely to adopt a reviewer-like form, because the retrieved historical review provides an explicit reference of how similar issues were previously critiqued.

Second, retrieval can affect \emph{what} the model chooses to comment on.
In this example, the reference flags \emph{unnecessary synchronization}.
When retrieval returns a historical critique that points to synchronization redundancy, the generated output becomes more likely to discuss that specific issue rather than producing a generic rewrite or a broad explanation.

This example also highlights a known limitation of n-gram based metrics: they can reward short outputs with incidental overlap and undervalue semantically appropriate but differently phrased critiques.
We therefore include human evaluation next.

\begin{table}[t]
\centering
\small
\setlength{\tabcolsep}{2.3pt}
\renewcommand{\arraystretch}{1.05}
\caption{Human evaluation of RARe's generated reviews on 100 samples from the CRer. and Tuf. datasets.}
\label{tab:human_evaluation}
\begin{tabular*}{\columnwidth}{@{\extracolsep{\fill}}lcccccccc@{}}
\toprule
 & \multicolumn{4}{c}{CRer.} & \multicolumn{4}{c}{Tuf.} \\
\cmidrule(lr){2-5} \cmidrule(lr){6-9}
Metrics & DI & +RA & FT & +RA & DI & +RA & FT & +RA \\
\midrule
Perfect prediction & 0 & 0 & 0 & 0 & 0 & 0 & 0 & \textbf{3} \\
Semantically equivalent & 0 & 1 & 1 & 4 & 1 & 3 & 7 & 11 \\
Alternative solution & 12 & 44 & 55 & 64 & 14 & 52 & 52 & 50 \\
Incorrect & 88 & 55 & 44 & 32 & 85 & 45 & 41 & 36 \\
\bottomrule
\end{tabular*}
\end{table}
\subsubsection{Human evaluation}
\label{sec:human_eval}

Following prior work~\cite{tufano2022using,hong2022commentfinder}, the quality of the code reviews can be evaluated manually by four metrics:

\textbf{Perfect prediction:} the generated review is effectively equivalent to the reference. 
\textbf{Semantically equivalent:} the generated review expresses the same critique as the reference but with different wording.
\textbf{Alternative solution:} the generated review differs from the reference but is still a useful and reasonable critique for the code.
\textbf{Incorrect:} the review is not useful for the code (irrelevant, generic, or misleading).

We randomly sample 100 instances from the test split of each dataset and evaluate outputs from Llama~3.1 (our strongest generator).
Two expert annotators (6+ years of software engineering experience) label each output independently and resolve disagreements through discussion.
We treat \emph{Perfect prediction}, \emph{Semantically equivalent}, and \emph{Alternative solution} as valuable reviews.

\begin{figure*}[t]
    \centering
    \includegraphics[width=\textwidth]{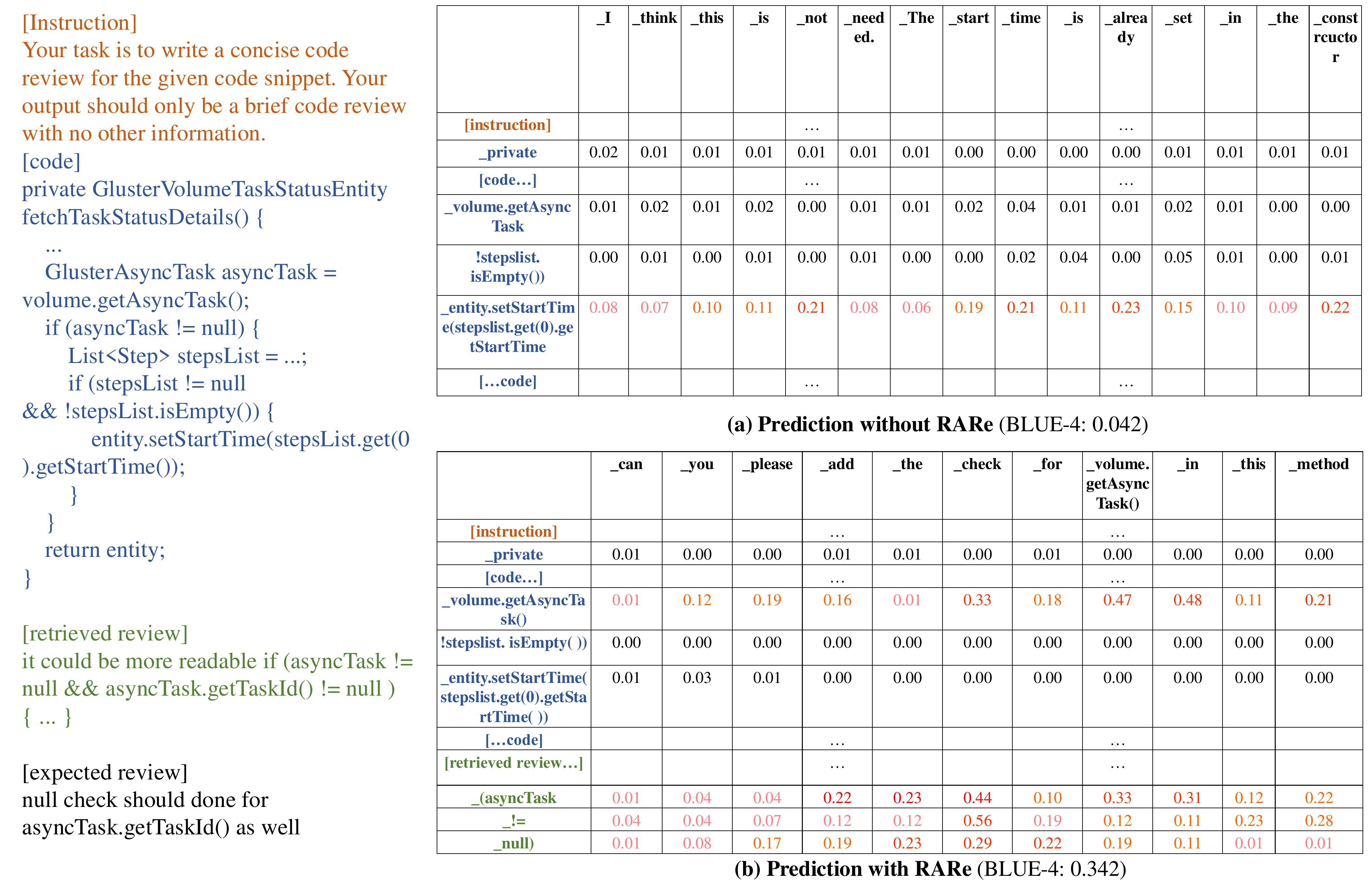}
    \caption{An example from Tuf. dataset and the saliency heatmap comparison for fine-tuned Llama-3.1-8B with and without retrieval augmentation. The horizontal rows of tables display the models' output. Different input components are differentiated using colors, and higher attention scores are highlighted in red within the table.}
    \label{fig:case-study}
\end{figure*}

Table~\ref{tab:human_evaluation} confirms that retrieval augmentation substantially increases the fraction of valuable reviews, especially under direct inference.
On Tuf., valuable reviews rise from 15 (DI) to 55 (DI+RA), and on CRer.\ from 12 (DI) to 45 (DI+RA).
After fine-tuning, retrieval remains helpful: on CRer.\ the number of valuable reviews increases further, and on Tuf.\ retrieval increases the number of high-agreement cases (Perfect prediction and Semantically equivalent), consistent with better alignment to the intended critique.

\subsection{Interpretative Analysis of RARe}
\label{inseq}
We use \emph{Inseq}, an interpretability toolkit for sequence generation models, to visualize token-level attribution between the input and the generated output (code and retrieved review). It produces attention-based saliency scores that indicate which input tokens the model relies on when generating each output token.

As depicted in Figure~\ref{fig:case-study}, the performance of RARe is notably superior, as partly evidenced by its higher BLEU-4 score (0.342 v.s. 0.042). Specifically, the text retrieved (marked in green) receives considerable focus in RARe's predictions, as indicated by its high attention score. This demonstrates that the model effectively leverages external information from retrieval, which is a suggestion to "incorporate a null check function" in this example. Moreover, RARe doesn't simply copy the retrieved results, it also concentrates on the code. This is evident from the attention scores assigned to the code (highlighted in blue) from the output in Table (b), showing RARe conducts a balanced integration of external knowledge with original code analysis.

Going back to a broader perspective, it becomes apparent that retrieval results can significantly enhance the model's attention, preventing it from concentrating on less relevant code. For instance, without RAG, the model focus on the \emph{getStartTime()} function, diverging from our expected review. However, with RAG, this deviation is corrected, redirecting the model's attention to the null check in \emph{volume.getAsyncTask()} function. It is important to clarify that the output in Table (a) is not necessarily incorrect; rather, it deviates from the reviewer's intended focus. This deviation suggests that the output lacks specific domain knowledge necessary for concentrating on the aspects of the code that truly require review.

\section{Discussion}
\label{sec:discussion}

Our findings suggest that retrieval augmentation for code review is best understood as a task-adaptive RAG setting, where retrieved historical reviews mainly serve as in-context examples that shape how the model writes and what it chooses to comment on.
This view yields two implications.

First, more retrieval is not always better.
Across both datasets, top-1 retrieval consistently performs best, while adding more retrieved items (top-3/top-5) degrades performance.
This indicates that historical reviews encode not only lexical cues but also critique direction and tone; combining multiple reviews can introduce redundancy or conflicting cues that distract the generator when prompt length is limited.
In practice, retrieval for code review should emphasize careful selection and redundancy control rather than simply increasing the number of retrieved items.
This observation is also consistent with recent work on selective retrieval and context compression~\cite{hwang2025exit,louis2025pisco,jin2025sara}.

Second, retriever choice should match corpus characteristics.
We observe that code--review alignment retrieval (DPR-style) is more robust on multi-language corpora, whereas code-only similarity with reranking remains competitive on single-language corpora.
This suggests that the most suitable retrieval signal depends on how reliable code similarity is as a proxy for review relevance in the target setting.

\section{Conclusion}
\label{sec:conclusion}

We presented RARe, a retrieval-augmented framework for code review generation that prepends retrieved historical reviews to the prompt as in-context examples.
Across two benchmarks, RARe outperforms strong baselines and yields especially large gains for direct inference.
A central finding is that adding more retrieved examples can hurt: using only the top-1 retrieved example is most effective, while top-3/top-5 often degrade performance due to redundancy and conflicting cues in historical reviews.
Human evaluation further supports that retrieval-augmented prompting reduces generic outputs and improves review focus.
This study is limited to two public benchmarks and a fixed set of retrievers and prompting choices.
Future work will explore hybrid retrieval, learned reranking for intent-aligned selection, and redundancy-aware context construction when multiple retrieved candidates are available.

\section*{Acknowledgment}
This work was funded by the China Scholarship Council (CSC) and supported by the Leiden Institute of Advanced Computer Science (LIACS) and SURF.
\clearpage

\bibliographystyle{plain}
\bibliography{rag_fse}

\end{document}